\documentstyle[twocolumn,aps,epsfig,here]{revtex}

\begin{document}
\title{
Event-by-event fluctuations of average transverse momentum 
in central Pb+Pb  \\ collisions at 158~GeV per Nucleon 
}
\author{
H.~Appelsh\"{a}user$^{7,\#}$, J.~B\"{a}chler$^{5}$,
S.J.~Bailey$^{17}$, D.~Barna$^{4}$, L.S.~Barnby$^{3}$,
J.~Bartke$^{6}$, R.A.~Barton$^{3}$, L.~Betev$^{12}$, H.~Bia{\l}\-kowska$^{15}$,
A.~Billmeier$^{10}$
% ...
\\
(NA49 Collaboration)
}

\address {
% ...
$^{3}$Birmingham University, Birmingham, England.\\
$^{4}$KFKI Research Institute for Particle and Nuclear Physics, Budapest, Hungary.\\
$^{5}$CERN, Geneva, Switzerland.\\
$^{6}$Institute of Nuclear Physics, Cracow, Poland.\\
$^{7}$Gesellschaft f\"{u}r Schwerionenforschung (GSI), Darmstadt, Germany.\\
% ...
$^{10}$Fachbereich Physik der Universit\"{a}t, Frankfurt, Germany.\\
% ...
$^{12}$University of California at Los Angeles, Los Angeles, USA.\\
% ...
$^{15}$Institute for Nuclear Studies, Warsaw, Poland.\\
% ...
$^{17}$Nuclear Physics Laboratory, University of Washington, Seattle, WA, USA.\\
% ...
$^{\#}$present address: Physikalisches Institut, Universit\"at Heidelberg, Germany\\
% ...
}

\date{\today}
\maketitle

\begin{abstract}\noindent
We present first data on event-by-event fluctuations in the average transverse momentum of 
charged particles produced in Pb+Pb collisions at the 
CERN SPS.
This measurement provides  previously unavailable information allowing sensitive 
tests of microscopic and thermodynamic collision models
and  to search for fluctuations expected to occur in the vicinity of the predicted
QCD phase transition.
We find that the observed variance of the event-by-event average transverse 
momentum is consistent with independent particle production modified by the 
known two-particle correlations due to quantum statistics and final state
interactions and folded with the resolution of the NA49 apparatus. 
For two specific models of non-statistical fluctuations in transverse momentum 
limits are derived in terms of fluctuation amplitude.
We show that a significant part of the parameter space 
for a model of isospin fluctuations predicted as a consequence of 
chiral symmetry restoration in a non-equilibrium scenario is excluded by our measurement.
\end{abstract}
\section{Introduction}
The ultimate goal in the study of collisions of heavy ions at the 
highest available energies is the production and characterization of an 
extended volume of deconfined quarks and gluons, the quark gluon plasma
(QGP)\cite{qm97}.
Due to the high multiplicities in central Pb+Pb collisions at 158~GeV per Nucleon, 
recorded in the NA49 large acceptance spectrometer, 
a statistically significant determination
of momentum space distributions and particle ratios can be performed
for single events, allowing for a study of 
event-by-event fluctuations\cite{sto95,rol97}. In this paper we will 
focus on fluctuations in the average transverse momentum of individually measured 
charged particles from event to event.\\
One expects that the fluctuation patterns
are altered in the vicinity of the QCD phase transition \cite{sto95,mrow93}.
This conjecture is supported by recent calculations in an effective 
model of the strong interaction \cite{step98,stephanov99}, which suggest
that near a tri-critical point in the QCD phase diagram the event-by-event fluctuation 
pattern in transverse momentum should change significantly. \\
 A precise measurement of event-by-event fluctuations allows for a 
test of the hypothesis of thermal equilibrium \cite{gazd92} 
and the extraction of thermodynamical properties of the system in a model comparison.
Model studies\cite{random} have shown that 
non-equilibrium models of nuclear collisions based purely on 
initial state scattering can be tested by measurements of 
transverse momentum fluctuations. Model calculations on transverse
momentum fluctuations have been performed in many of the commonly
used microscopic models of nuclear collisions \cite{bleicher98,liu98,cap99},
in particular focussing on the question of how the fluctuations
change when going from nucleon-nucleon to nucleus-nucleus collisions.\\
It has also been suggested that for a thermodynamical picture of the 
strongly interacting system formed in the collision, the strength of 
fluctuations is directly related to fundamental properties of the 
system like the specific heat \cite{stod95,shur98a}
and  matter compressibility \cite{mrow98a}. A detailed discussion
of transverse momentum fluctuations in a resonance gas model can 
be found in \cite{stephanov99}.\\
One of the most intensely discussed topics related to fluctuations
at the QCD phase transition is the formation of so-called disoriented 
chiral condensates (DCCs) \cite{raja93,anselm91} as a consequence of 
the transient restoration of chiral symmetry, which may lead to a production of 
pions with much larger fluctuations of the charged-to-neutral pion ratio than
expected from Poisson-statistics.
The sensitivity of our  measurement to these 
fluctuations is discussed.\\
NA49 is currently pursuing two different,
but complementary approaches to the characterization of the 
single events. In the approach presented here
we characterize the event by global observables 
like the mean transverse momentum of individually detected charged particles in the event,
averaging over a large interval in momentum space.
Global quantities in general also include contributions
from particle correlations occuring at smaller scales, i.e.
smaller intervals in momentum space.
NA49 is also studying a system of differential measures of 
event morphology which aim at a multiscale characterization
of the correlation content of single events, which should eventually
provide a decomposition of the global fluctuations as a 
function of scale \cite{train98b}.
\section{Experimental Setup and Data selection}
The setup of the NA49 experiment is described in \cite{nimpaper}.
We used a data set of central Pb+Pb collisions that where   
selected by a trigger on the energy deposited in 
the NA49 forward calorimeter. The trigger accepted only the 5\% most 
central events, corresponding to an impact parameter range of $b < 3.5$~fm. 
The event vertex was reconstructed using information from beam position
detectors and the fit of the measured particle trajectories. Only events
uniquely reconstructed at the known target position were used.
The NA49 large acceptance hadron spectrometer allows the detection of 
more than 1000 individual charged particles for a single central Pb+Pb
collision.\\
In this analysis particles were selected that had a measured track length of 
more than 2~m in one of the two  Main Time Projection Chambers (MTPC) outside 
the magnectic field and were also observed in at least 
one of the Vertex TPC's inside the superconducting magnets.
We studied particles in a region of $0.005 < p_T < 1.5$~GeV/c 
and rapidity $4 < y_{\pi} < 5.5$. A cut on the extrapolated impact parameter 
of the particle track at the primary vertex 
was used to reduce the contribution
of non-vertex particles originating from weak decays and secondary
interactions. We estimate that about 60~\% of such particles are 
rejected by the vertex cuts.
From a full simulation of our apparatus using a GEANT \cite{geant} based 
Monte-Carlo code and a parametrization of the detector response 
we obtained an average reconstruction efficiency of 90\%. The
average resolution in transverse momentum for the particles used here 
is around 3~MeV/c, dominated by multiple Coulomb scattering.
The two-track resolution was determined using both the 
simulation and a mixed-event technique. For particles selected 
by the track cuts the pair 
detection efficiency drops from around 80\% at an average distance in
the Main TPC of $d = 2.5$~cm to around 20\% at an average distance
of $d = 1.5$~cm. 
In table~1 the most important parameters of the inclusive
and event-by-event distributions of accepted particles are summarized.
Throughout this paper brackets ($\langle x \rangle$) will
denote averages over events and bars ($\overline{x}$) will denote inclusive averages over
all (accepted) particles and all events.
\begin{table}[t]
\caption{Measured parameters of the inclusive and event-by-event accepted particle distributions.
Errors are statistical only.}
\begin{tabular}{|l|l|}
No. of events & 98426 \\ 
\tableline
$\langle N\rangle$ & $ 270.13 \pm 0.07 $ \\ 
\tableline
$(\langle N^2\rangle - \langle N\rangle^2)^\frac{1}{2}$ &$23.29 \pm 0.05 $ \\
\tableline
$\overline{p_T}$ & $376.75 \pm 0.06$~MeV/c \\
\tableline
$(\overline{p_T^2} - \overline{p_T}^2)^\frac{1}{2}$ & $282.2 \pm 0.1$~MeV/c \\
\end{tabular}
\end{table}
\section{Analysis and results}
For each of the events we characterize the observed particle distribution in the 
acceptance region by calculating the mean of the transverse momentum 
distribution of the $N$ accepted particles in the event,
\begin{equation}
M(p_T) = 1/N \cdot \sum_{i=1}^{N} p_{Ti}.
\end{equation}
The resulting distribution of $M(p_T)$ is shown in fig.~1. \\
The distribution of $M(p_T)$ has approximately Gaussian shape. 
No significant excess of 'anomalous' events outside the main distribution
is observed.
For the variance of 
the $M(p_T)$ distribution we get
\begin{eqnarray*}
 V(M(p_T))/\overline{p_T} = 4.65 \pm 0.01\%.
\end{eqnarray*}
The biggest contribution to the observed variance is expected to come from 
finite-number statistics. The main task in the remainder of the paper will
be to extract possible non-statistical contributions
on top of the trivial statistical variation from event to event.
A first impression of the possible size of non-statistical contributions can
be obtained by a comparison to the same distribution calculated for 
so-called mixed events 
(solid line in fig.~1). The mixed events were constructed by combining 
particles drawn randomly from different events while reproducing 
the multiplicity distribution of the real events.  
\noindent Only one track 
of any original event was used in a given mixed event and no further selection
was made regarding the impact parameter or multiplicity of the 
original events.
By construction the mixed events have the same single-particle distributions as the real
events, but no internal correlations. The variance of the mixed event $M(p_T)$ 
distribution is therefore determined by finite number statistics, giving
\begin{eqnarray*}
(\overline{p_T^2} - \overline{p_T}^2)^\frac{1}{2}/
( \overline{p_T} \cdot \sqrt{\langle N \rangle}) = 4.6\%.
\end{eqnarray*}
The mixed event distribution resembles, very closely, the single event distribution,
thus suggesting that large amplitude non-statistical fluctuations are small and/or
rare.\\
To further quantify and study the deviation of the $M(p_T)$ distribution
a number of methods has been discussed recently \cite{kadija92,bialas99,alberico99,belkacem99}.
In this analysis we follow the approach suggested in \cite{gazd92}.
We define for every particle $i$
\begin{equation}
z_i = p_{Ti} - \overline{p_T} .
\end{equation}
For every event we calculate
\begin{equation}
Z = \sum_{i=1}^{N} z_i .
\end{equation}
With this definitions
we use the following measure to quantify the degree of fluctuation
in mean transverse momentum from event to event:
\begin{equation}
\Phi_{p_T}  = \sqrt{\frac{\langle Z^2 \rangle}{\langle N \rangle}} - \sqrt{\overline{z^2}}.
\end{equation}
One limiting case for this fluctuation measurement is particle emission according 
to a parent distribution that 
remains unchanged for all events, i.e.\
every single event is just a 
random sample of finite multiplicity taken from the same parent 
distribution. $\Phi_{p_T}$ was defined such that for this case a 
value of zero is assumed. This value also corresponds to the fluctuations
for an ideal gas of classical particles\cite{mrow98b}. \\
\begin{figure}[t]
\centerline{\epsfig{file=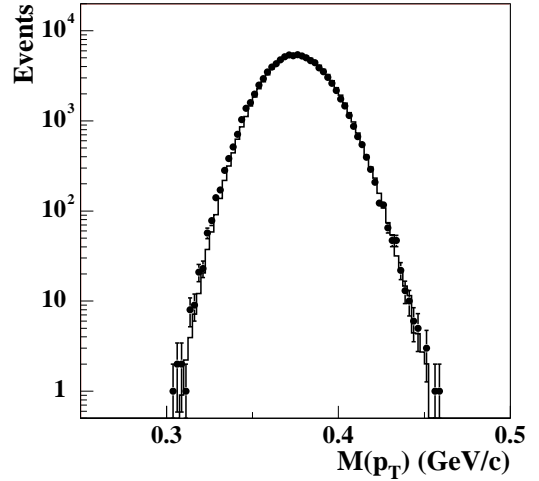,height=7.5cm }}
\caption{Event-by-event distribution of the mean transverse 
momentum $M(p_T)$ of accepted particles in the event (points). For
comparison, the solid line shows the $M(p_T)$ distribution
for mixed events.}
\label{1}
\end{figure}
Our goal is to detect or exclude fluctuations that are 
compatible with changes, event-to-event, in the 
parent distribution in transverse momentum.
Such changes would in general lead 
to values of  $\Phi_{p_T} > 0$. $\Phi_{p_T}$, as we will demonstrate
later, is also sensitive to internal correlations or anti-correlations
of particles within single events, which result in  $\Phi_{p_T} > 0$
or  $\Phi_{p_T} < 0$, respectively.\\
In our data set we measure a value of 
\begin{eqnarray*}
\Phi_{p_T} = 0.6 \pm 1.0~\mbox{MeV/c},
\end{eqnarray*}
compatible with zero.
The error was estimated by calculating $\Phi_{p_T}$ separately
on independent subsamples of approximately 10000 events each.
Before discussing the implications of the small value measured
for $\Phi_{p_T}$ for the existence of collective non-statistical 
fluctuations, we first turn to investigating the sensitivity of 
this measure to correlations at small scales, which are known to
be present in these events.

Previous studies \cite{hbtpaper} have quantitatively analyzed two-particle
correlations in relative momentum, with quantum statistics and 
Coulomb final state interactions giving the strongest contributions.
We have developed an analysis procedure to identify the contribution
of these known effects, folded with the NA49 experimental response,
to the observed value of  $\Phi_{p_T}$. 
This comparison is done as follows.
\begin{enumerate}
\item From the original events we construct a sample  
of mixed events with the same overall multiplicity 
distribution as the real events.
Without further modifications, the mixed events give a 
value of $\Phi_{p_T} = -0.2 \pm 0.4$~MeV/c, again consistent with zero.
\item In the second step we model 
the  contributions from particle pair
correlations at small relative momenta ('small scales'). 
The effect of Bose-Einstein or Fermi-Dirac
statistics on transverse momentum fluctuations has been discussed in 
\cite{mrow98b}. \\
The effects of quantum statistics and final-state (Coulomb) interactions 
partially cancel each other and are further diminished by particles emitted
from long-lived resonances, particles originating from weak decays 
and by including combinations of non-identical particles. 
A detailed evaluation of two-particle correlations in NA49 is 
given in \cite{hbtpaper}.
To include the sum of all these effects in the mixed events 
we use a procedure that was described in \cite{kad92}. In this procedure
the momenta of particles are altered pairwise to introduce the desired
form of the two-particle correlation function, making sure
that in the mixed events on the average the two-particle correlations as a function 
of relative momentum 
closely match those
observed in the data, only corrected for the two-track resolution. 
The analysis of the modified mixed events  provides an estimate of the 
minimal event-by-event $M(p_T)$ fluctuations that we expect 
as a consequence of the observed average two-particle correlations.
The contribution from the two-particle correlation function 
alone is $\Delta \Phi_{p_T} = 5 \pm 1.5$~MeV/c. 
\item In our data set we observe a slight but statistically significant
correlation between 
the multiplicity of the event $N$ and the average transverse momentum
$M(p_T)$. The correlation is characterised by a linear correlation 
coefficient of 
\begin{eqnarray*}
\frac{\langle (M(p_T) - \overline{p_T}) \cdot (N - \langle N ) \rangle}
{V(M(p_T)) \cdot V(N)} = -0.03 \pm 0.01,
\end{eqnarray*} where $V(N)$ is the variance of the multiplicty distribution. We therefore
observe a slight decrease of $M(p_T)$
with increasing multiplicity within the central collision data set. 
Introducing this correlation in the 
mixed events gives a negligible contribution to the width of the 
$M(p_T)$ distribution ($\Delta \Phi_{p_T} \ll 1$~MeV/c).
\item We then apply an experimental filter on each of the modified mixed events 
that simulates the influence of the two-track resolution and momentum
resolution of the NA49 apparatus. While the contribution from
momentum resolution is found to be negligible for the range of 
fluctuations considered here, the two-track resolution results in an 
effective anti-correlation between particles in momentum space
and gives a contribution of $\Delta \Phi_{p_T}  = -4 \pm 0.5$~MeV/c.
\end{enumerate}
Combining all effects, we find that the observed value $\Phi_{p_T}$ is
compatible with independent particle production:
Including both the effects of two-particle correlations in momentum 
space and the experimental two-track resolution leads to a 
cancellation resulting in a very small net contribution.
Any additional contributions beyond those mentioned above either have 
to be small or cancel with sufficient accuracy to be compatible 
with $\Phi_{p_T} = 0.6$~MeV/c. \\
Here it is worthwhile to note that the effects of two-particle 
momentum correlations and two-track resolution, as included 
in our simulations, both are strongly multiplicity-dependent
and become negligible for multiplicities comparable to those 
observed in p+p collisions. This suggests that the physical origin of
transverse momentum fluctuations as measured using $\Phi_{p_T}$
changes when comparing the value of 0.6~MeV/c for Pb+Pb to the 
preliminary NA49 measurement of $5\pm 1$~MeV/c for p+p collisions
\cite{roland98}.

To further study the sensitivity of our
measurement we have introduced explicit non-statistical fluctuations
in the mixed event sample and studied the response in $\Phi_{p_T}$ 
as a function of the parameters controlling the strength of these 
fluctuations. By comparing the value obtained for
$\Phi_{p_T}$ observed in such models with that in the data, we 
can determine the sensitivity of our measurement to various 
kinds of fluctuations and eventually derive limits on the 
amplitude or frequency of occurence for fluctuations in specific 
models.\\ 
In the first model we examine the sensitivity to non-statistical 
fluctuations introduced by scaling the transverse momentum for 
all tracks in a given event by a constant factor $x$. The resulting
change in the $p_T$ parent distribution from event to event resembles
that of an event by event change in the inverse slope parameter of 
an exponential transverse momentum distribution. 
We obtain a random number $x$ for each event distributed
according to 
\begin{equation}
P(x) = \frac{1}{\sqrt(2 \pi \sigma_{fluc}^2)} \exp(-(x-1)^2/(2 \sigma_{fluc}^2)) 
\end{equation}
and multiply the  transverse momentum of each particle in the event  by
the same factor $x$. Here the amplitude of the event by event fluctuation
is controlled  by the parameter $\sigma_{fluc}$.
Adding these fluctuations to the mixed events and neglecting
two-particle correlations and two-track resolution, $\Phi_{p_T}$ increases 
proportionally to $\sigma_{fluc}^2$.\\
When adding all effects, we find that for this model a fluctuation strength 
$\sigma_{fluc} = 1.2$\% corresponds to $\Phi_{p_T} = 7$~MeV/c. Given 
the definition of $\Phi_{p_T}$ in Eq.~4 and net contributions to $\Phi_{p_T}$ as 
observed and simulated in steps 
1 to 4 above we can establish an upper limit on $\sigma_{fluc} < 1.2\%$ at 
90\% confidence level.\\
While a precise limit for specific models can only be set using a simulation
procedure as outlined above, we typically find that for various types
of non-statistical fluctuations our measurement is sensitive when the 
fluctuations lead to an effective non-statistical variation in the mean 
transverse momentum from event to event of about 1\%.

We can also use the simulation procedure to establish limits on fluctuations of amplitude 
$\sigma_{fluc}$ that occur only in a fraction $F$ of all events. 
The resulting exclusion plot is shown in fig.~2, 
where the relative frequency $F$ of events exhibiting  
fluctuations of amplitude $\sigma_{fluc}$ is plotted versus
$\sigma_{fluc}$.
We see that for $F = 1$ fluctuations of a relative
amplitude of $\sigma_{fluc} > 1.2\%$ are ruled out at 90\% confidence
level, whereas fluctuations occuring in 1\% of the events
can only be ruled out for $\sigma_{fluc} > 10\%$.
\begin{figure}[htp]
\centerline{\epsfig{file=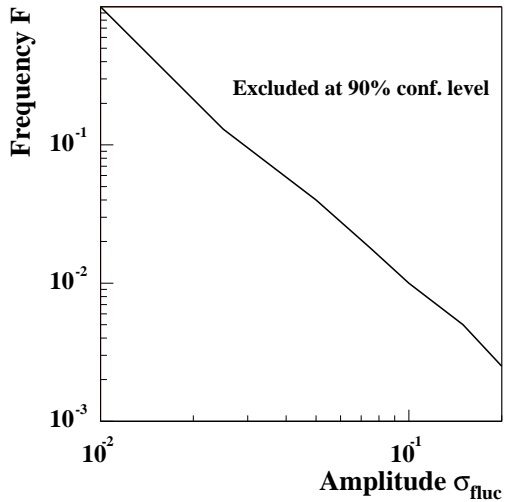,height=7.5cm}}
\caption{Limit on the amplitude of fluctuations in the 
$p_T$ parent distribution
as a function of the frequency of events showing the fluctuation.}
\label{3}
\end{figure}

It is important to note that the measurement of fluctuations in 
$M(p_T)$ is also
relevant for models of processes that lead to 
non-statistical fluctuations \em localized \em in transverse
momentum. The most widely discussed example of such
a process is the formation of
disoriented chiral condensates (DCCs)\cite{raja93,anselm91}, which has been
postulated as a consequence of the possible restoration
of chiral symmetry in non-equilibrium scenarios for heavy-ion collisions.
DCC models predict the formation of domains 
that eventually emit pions where the ratio $f$ of neutral 
to all pions varies as 
\begin{equation}
P(f) = \frac{1}{2 \sqrt{f}}.
\label{dcc}
\end{equation}
The models also suggest that pions emitted from DCC domains
will be preferentially produced at low transverse
momenta\cite{raja93}. 
This provides for a translation of the number-fluctuations
predicted by the DCC models into $p_T$ fluctuations accessible
to our experiment. A limit on DCC production has already been
set by the WA98 collaboration \cite{agga98}, based on a study of 
fluctuations in the relative multiplicties of charged and neutral 
particles near mid-rapidity.\\
For comparison purposes we used the same DCC model as in  \cite{agga98},
where the DCC production is characterized by the probability $F$
to form a single DCC domain in an event and the fraction $\xi$
of pions coming from the DCC. We make the additional assumption
that the DCC pions are produced with $p_T < p_T^{max} =  m_{\pi}$.
The ratio of neutral to charged
pions was chosen randomly according to equ.~\ref{dcc}. 
The isospin fluctuations of pion production from DCCs then
lead to multiplicity fluctuations of charged pions at low
transverse momenta and  therefore to non-statistical fluctuations in $M(p_T)$.
For DCCs occuring in every event ($F = 1$) the fluctuations observed in the 
data rule out DCC sizes
of $\xi > 3.5$~\%, which is about a factor of 5 smaller than 
the previous limit set in \cite{agga98}. This limit could be further
improved by restricting the analysis to the region of  small
transverse momenta.\\
\section{Summary}
In summary, event-by-event fluctuations  in the average transverse 
momentum of charged particles in the forward hemisphere
of central Pb+Pb collisions have been measured. 
The distribution of average transverse momentum 
per event $M(p_T)$ has an approximately Gaussian shape, with
no excess of 'anomalous' events falling out of the distribution.
The fluctuation strength in the data is characterized by
a value of $\Phi_{p_T} = 0.6 \pm 1$~MeV/c.
Using a procedure based on mixed events
we find that the fluctuations in $M(p_T)$ from event to event
are compatible with 
independent particle production modified by the 
known two-particle correlations due to quantum statistics and final state
interactions and  taking into account the response
of the NA49 apparatus, without requiring further variations in the 
transverse momentum parent distribution from event to 
event.\\
For a model of non-statistical fluctuations in average $p_T$ we
use a detailed simulation procedure to determine an upper limit on 
the strength of fluctuations occuring in every event of 
$\sigma_{fluc} < 1.2\%$ at 90\% confidence level.\\
We also demonstrate that high precision measurements of charged particle
transverse momentum fluctuations provide a sensitive test for models
predicting the formation of disoriented chiral condensates in heavy-ion collisions.
Finally, we have provided the first measurement to compare to predictions of 
event-by-event transverse momentum fluctuations in thermodynamical descriptions of the strongly 
interacting system produced in Pb+Pb collisions at the SPS.\\

This work was supported by the Director, Office of Energy Research,
Division of Nuclear Physics of the Office of High Energy and Nuclear Physics
of the US Department of Energy under Contracts DE-ACO3-76SFOOO98 and DE-FG02-91ER40609,
the US National Science Foundation,
the Bundesministerium fur Bildung und Forschung, Germany,
the Alexander von Humboldt Foundation,
the UK Engineering and Physical Sciences Research Council,
the Polish State Committee for Scientific Research (2 P03B 02615 and 09913),
the Hungarian Scientific Research Foundation under contracts T14920 and T23790,
the EC Marie Curie Foundation,
and the Polish-German Foundation.

\end{document}